\documentclass[floatfix,superscriptaddress,twocolumn]{revtex4}
\usepackage{graphicx}
\usepackage{latexsym}
\usepackage{amsmath, amsthm, amssymb}
\usepackage{epsfig}

\begin{document}

\title{Kramers problem for a dimer: effect of noise correlations}
\author{R. K. Singh}
\email[]{rksingh@imsc.res.in}
\affiliation{The Institute of Mathematical Sciences, 4th Cross Road, CIT Campus,
Taramani, Chennai- 600113, India}

\begin{abstract}
Kramers problem for a dimer in a bistable piecewise linear potential is studied in 
the presence of correlated noise processes. The distribution of first passage times 
from one minima to the basin of attraction of the other minima is found to have 
exponentially decaying tails with the parameter dependent on the amount of correlation 
and the coupling between the particles. Strong coupling limit of the problem is analyzed 
using adiabatic elimination, where it is found that the initial probability density 
relaxes towards stationary value on the same time scale as the mean escape time. 
The implications towards polymer dynamics in a potential are discussed. 
\end{abstract}

\maketitle

\section{Introduction}
Escape of a particle confined in a metastable state is a ubiquitous problem arising 
in domains varying from chemical kinetics to transport theory. The theory of Brownian 
motion provides one of the most elegant approaches to study the problem by identifying 
the additional degrees of freedom as noise and friction \cite{hanggi}. This approach 
towards the escape problem was grounded in the seminal work of Kramers 
\cite{kramers, talkner}, who provided theoretical estimates for the rate of escape 
for a particle trapped in a metastable state in the limits of low and high friction. 

Generalizing the single particle problem, thermally activated escape of extended objects 
like polymers has attracted attention in recent times \cite{mesfin, park, lee, sebastian}. 
The studies have concluded that the escape of 
a polymer chain from a potential well depends nontrivially on the structural 
properties of the polymer, viz. the number of monomers constituting the chain and 
the strength of inter-particle interactions. The results provide a 
handle to control the rates of chemical reactions involving polymers by varying their 
structural parameters. Recently the problem of a dimer crossing a potential barrier 
has been investigated \cite{asfaw} conforming with the previous results for polymers. 
However, all these studies have focused on uncorrelated noise processes, whereas 
it is known that noises from identical origin are generally correlated \cite{fulinski}. 
Such correlations significantly affect the dynamics of a particle in a bistable potential 
\cite{jia, li, mei, xie}, and are known to induce nonzero transport in periodic 
potentials due to symmetry breaking \cite{sang}. The observations motivate us to 
study the effect of noise correlations on the dynamics of extended objects. 

In this paper we study the dynamics of the simplest extended object, a dimer: 
two harmonically coupled Brownian particles in a piecewise linear bistable potential. 
Additional thermal degrees of freedom are Gaussian white 
and correlated with each other. It is found that positively correlated noise processes 
facilitate barrier crossing for the dimer whereas negative correlations tend to diminish 
the effect of thermal degrees of freedom. The structure of the paper is as follows: 
in the next section the effect of coupling and correlation are studied on the motion 
of the dimer. Following it the strong coupling limit of the dynamics is analyzed and 
the effects of periodic forcing are also reported. The results are generalized 
to the dynamics of a polymer in a potential field with conclusions in the final 
section. 

\section{Dynamical system}
Let us start with the dynamical equations for a dimer in a bistable potential $U$: 
\begin{subequations}
\label{dyn}
\begin{align}
  \dot{x}_1 &= -U'(x_1) + F_{12}(x_1, x_2) + \eta_1(t),\\
  \dot{x}_2 &= -U'(x_2) + F_{21}(x_1, x_2) + \eta_2(t),
\end{align}
\end{subequations}
where $\eta_1$ and $\eta_2$ are Gaussian white noises of mean zero and correlations: 
\begin{subequations}
\label{corr}
\begin{align}
  \langle \eta_1(t) \eta_1(t') \rangle & = \langle \eta_2(t') \eta_2(t) \rangle 
  = 2 D \delta(t-t'),\\
  \langle \eta_1(t) \eta_2(t') \rangle & = \langle \eta_1(t') \eta_2(t) \rangle
  = 2 D \rho \delta(t-t'), 
\end{align}
\end{subequations}
with $D$ being the noise intensity and $\rho \in [-1,1]$ the measure of correlation. 
The noise intensity is a measure of the dimensionless temperature of 
the associated heat bath. Consequently, the existence of a correlation between the two 
noise processes is natural as $\eta_1$ and $\eta_2$ have the same thermal origin. 
The potential $U$ in eqn(\ref{dyn}) is a piecewise linear function defined as: 
\begin{align}
\label{potential}
U(x) = 
\begin{cases}
-x-1 ~~ \text{for} ~~ x < -1,\\
x+1 ~~ \text{for} ~~ -1 \leq x \leq 0,\\
-x+1 ~~ \text{for} ~~ 0 \leq x \leq 1,\\
x-1 ~~ \text{for} ~~ x > 1,
\end{cases}
\end{align}
having global minima at $x = \pm 1$ and a local maxima at $x = 0$. Components of 
the dimer interact via a harmonic potential $U_{sh} = \frac{k}{2}(x_1-x_2)^2$, 
with the corresponding forces $F_{ij}(x_1, x_2) = -\frac{\partial }{\partial x_i} 
U_{sh} (x_1, x_2)$ with $i \in \{1, 2\}$ and $i \neq j$, and $k$ being the spring 
constant. It is noted that the natural length of the spring is chosen to be negligibly 
small as compared to the separation of the global minima of the potential $U$ 
and hence, is ignored in the definition of the interaction potential $U_{sh}$. 

In order to diagonalize the correlation matrix in (\ref{corr}), let us transform 
the dynamical equations to $x_c = \frac{x_1+x_2}{2}$ and $x_r = \frac{x_1-x_2}{2}$, 
which are respectively the coordinates of the center of mass and relative separation 
between the two particles. In terms of the variables $x_c$ and $x_r$, the dynamical 
equations in (\ref{dyn}) are transformed as: 
\begin{subequations}
\label{dyn_cm}
\begin{align}
\label{cm1}
\dot{x}_c &= -\frac{U'(x_c+x_r)+U'(x_c-x_r)}{2} + \zeta_c(t),\\
\dot{x}_r &= -\frac{U'(x_c+x_r)-U'(x_c-x_r)}{2} - 2kx_r + \zeta_r(t),
\end{align}
\end{subequations}
where $\zeta_c = \frac{\eta_1+\eta_2}{2}$ and $\zeta_r = \frac{\eta_1-\eta_2}{2}$ 
are independent noise processes with mean zero and correlations: 
\begin{subequations}
\label{cm_corr}
\begin{align}
\label{corr1}
  \langle \zeta_c(t) \zeta_c(t') \rangle &= D (1+\rho) \delta(t-t'),\\
  \langle \zeta_r(t) \zeta_r(t') \rangle &= D (1-\rho) \delta(t-t'). 
\end{align}
\end{subequations}
The stochastic differential equations in (\ref{dyn_cm}) and (\ref{cm_corr}) are 
solved numerically using Heun's method \cite{toral} with the initial conditions 
$(x_c, x_r) = (-1.0, 0.02)$. 

\begin{figure}
\includegraphics[width=0.5\textwidth]{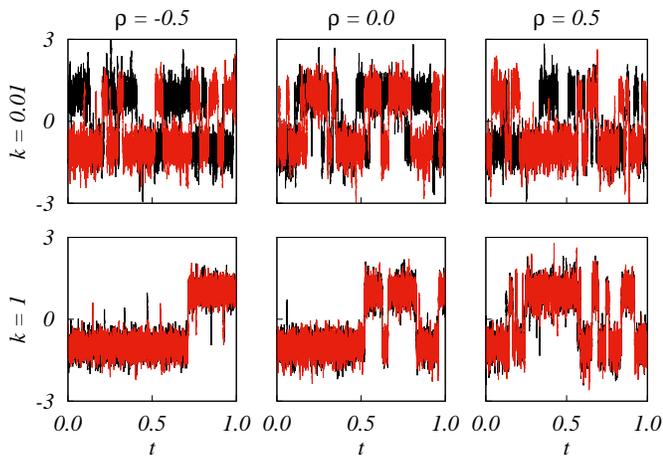}
\caption{Sample trajectories of the dimer $x_1$(black) and $x_2$(red) in the piecewise 
linear bistable potential $U$, with the particles interacting simple harmonically for 
the initial conditions $(x_c, x_r) = (-1.0, 0.02)$ for different values of spring 
constant $k$ and noise correlation $\rho$ for noise intensity $D = 0.25$. 
The time $t$ in the figure is in multiples of $10^3$. }
\label{fig1}
\end{figure}
Fig.~\ref{fig1} shows the trajectories of the dimer in the bistable potential 
$U$ for varying correlations $\rho$ and spring constant $k$ for noise intensity 
$D = 0.25$. The dependence of the nature of trajectories on the spring constant $k$ 
is evident from the figure. For low $k$ the two particles move nearly independent 
of each other, but for high $k$ the dimer moves as an effective single particle 
with the two particles fluctuating about the mean position independent of the value 
of noise correlation $\rho$. However, $\rho$ plays a decisive role in the dimer 
crossing the potential barrier when the coupling between the monomers is high, 
with positive correlation aiding in the back and forth hoping between the two 
minima and the negative $\rho$ confining the monomer in the stable position. 
To quantify the above observations let us study the residence time statistics 
of the center of mass in the potential wells, which identifies with the statistics of 
escape times \cite{choi}. 

\begin{figure}
\includegraphics[width=0.5\textwidth]{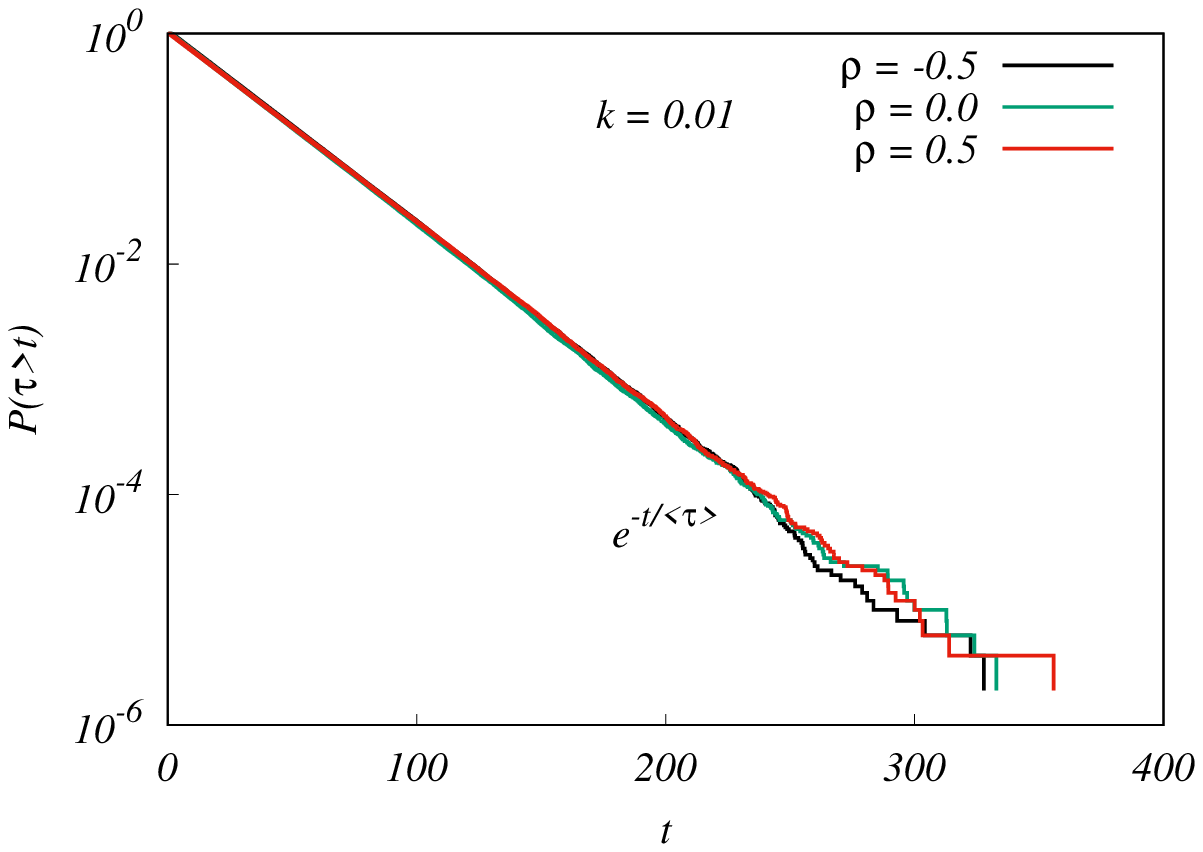}\\
\includegraphics[width=0.5\textwidth]{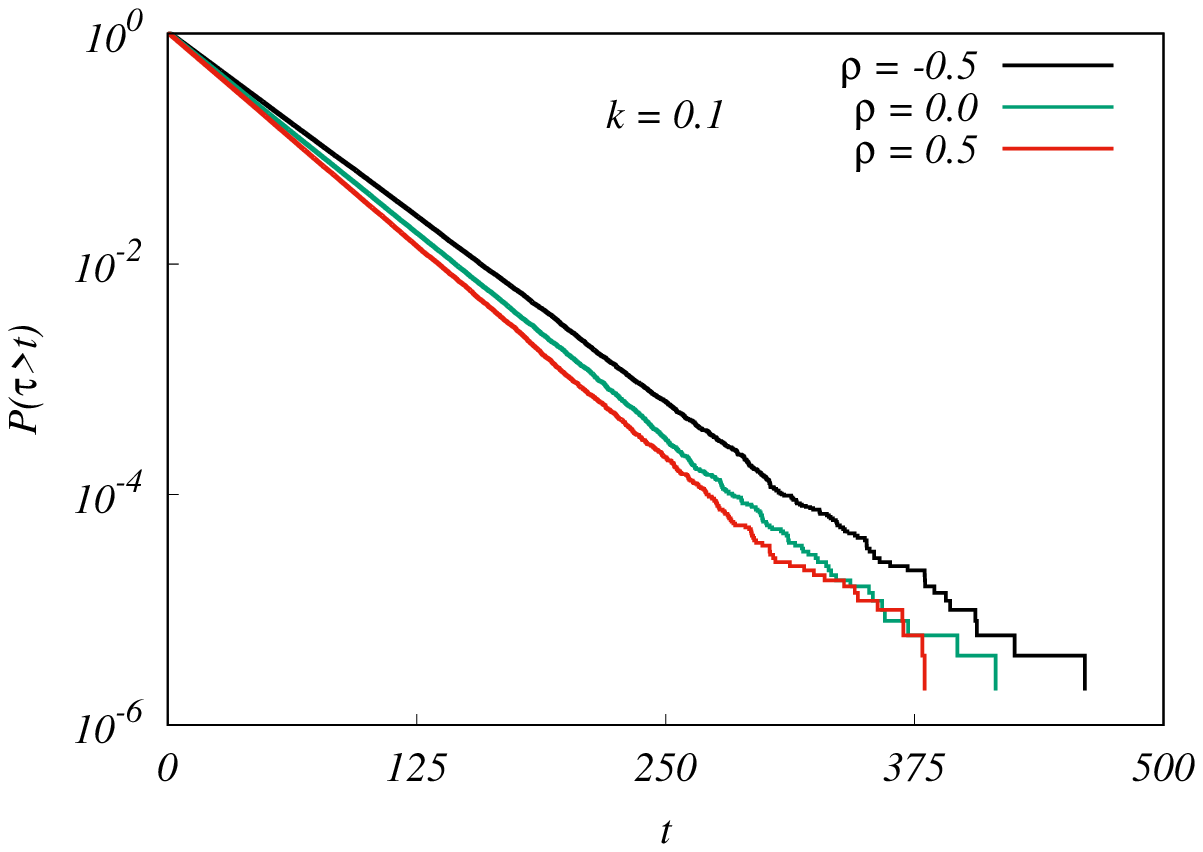}\\
\includegraphics[width=0.5\textwidth]{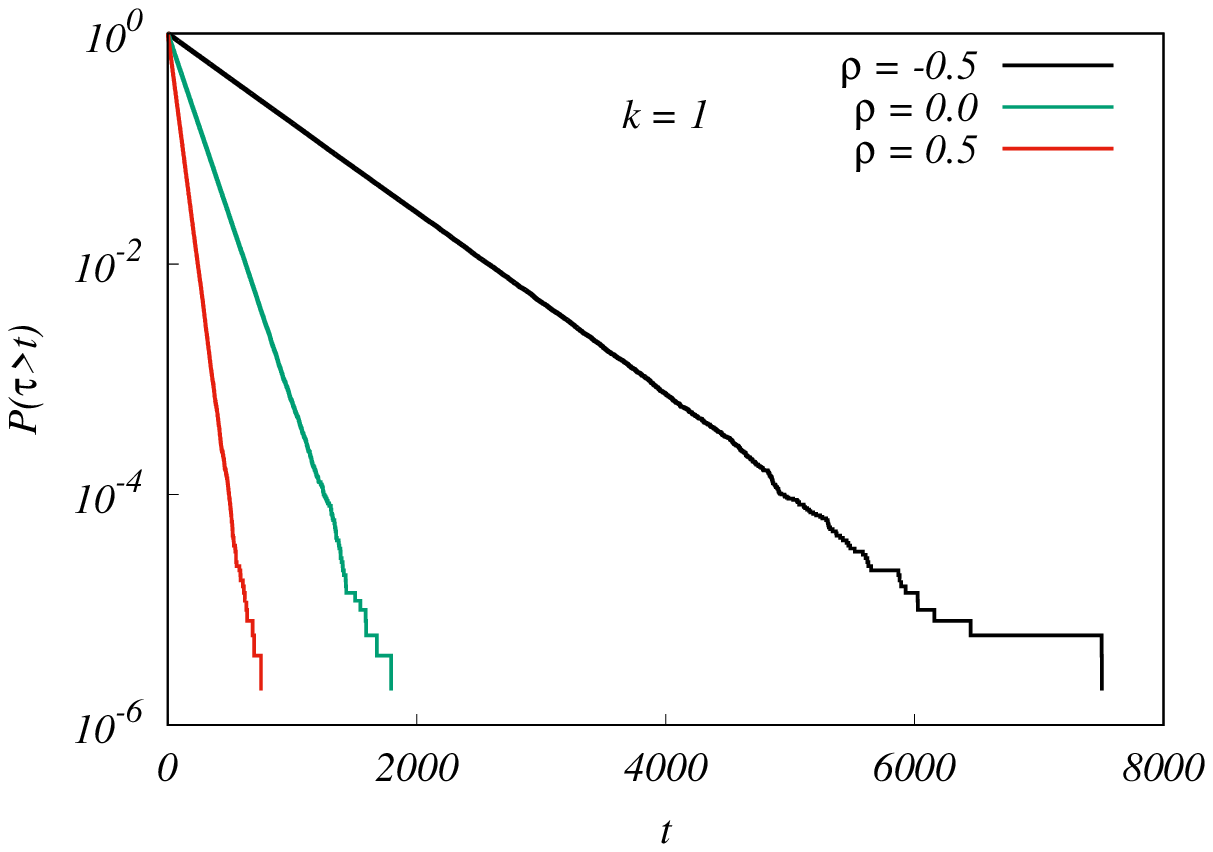}
\caption{Cumulative distribution of the first passage times $\tau$ of the center of mass 
starting in the left well to the absorbing boundary at $x_c = 0$. The distribution 
has exponentially decaying tails with parameter $\langle \tau \rangle$. The effect of 
the coupling between the two monomers is evident on the nature of the distribution: for 
nearly independent movement of the particles($k = 0.01$)(top) the distributions 
trace each other for different values of $\rho$. When the coupling between the monomers 
increases, the mean first passage time $\langle \tau \rangle$ is found to decrease with 
increasing $\rho$ as observed for $k = 0.1$(middle) and $k = 1$(bottom). 
The distributions are calculated using 500000 data points for noise 
intensity $D = 0.25$. }
\label{fig2}
\end{figure}
With the initial condition $x_c = -1$, lets look at the time it takes 
for the center of mass to reach the basin of attraction of the minima at $x_c = 1$. 
Fig.~\ref{fig2} shows the distribution of first passage times $\tau$ for different values 
of noise correlation $\rho$ and spring constant $k$ for noise intensity $D = 0.25$. 
The distribution shows exponentially decaying tails with parameter 
$\langle \tau \rangle$, the mean first passage time. However, when the coupling 
between the monomers is low($k = 0.01$), $\langle \tau \rangle$ is nearly independent 
of the correlation $\rho$. This is because for such low values of spring constant $k$, 
the particles move nearly independent of each other and hence the correlation between 
the thermal degrees of freedom does not have any significant impact on the rate of 
barrier crossing of the nearly independent particles. On the other hand, with increasing 
values of spring constant, e.g.- $k = 0.1 ~\text{and}~ 1$, it is observed that 
$\langle \tau \rangle$ decreases with increasing correlation $\rho$. 
The reason for such a behavior follows from the 
dynamical equations in (\ref{dyn_cm}) and (\ref{cm_corr}) which imply that the noise 
intensity affecting the dynamics of the center of mass is $D(1+\rho)$. As a result, 
for negative values of $\rho$ the center of mass does not feel the additive perturbations 
to the extent as felt in the absence of any correlations. Consequently, the escape to 
the absorbing boundary becomes difficult for negative values of $\rho$. 
On the other hand, $\rho > 0$ enhances the effect of the thermal degrees of freedom, 
making the transport of the dimer across the well relatively easier. The results 
detail us with the dynamical properties of coupled Brownian particles for different 
values of spring constant $k$ and noise correlation $\rho$. It is also inferred 
from the Fig.~\ref{fig2} that the mean escape time $\langle \tau \rangle$ is lowest when the 
two monomers move relatively independent of each other, i.e., for low values of spring 
constant $k$. The overall effect of coupling is to slow down the escape process and it 
is in this limit that the noise correlations play a significant role. Consequently, it 
becomes interesting 
to study the limit of large coupling constant in which the dimer moves effectively as 
a single particle at its center of mass and we proceed with this in the next section 
by the method of adiabatic elimination of the fast degrees of freedom \cite{risken}. 

\section{Adiabatic elimination}
The adiabatic elimination of the fast variable requires marginalization of the 
probability distribution $p(x_c, x_r, t)$  via the stationary solution of the 
Fokker-Planck operator for the fast variable $x_r$. The Fokker-Planck equation associated 
with the dynamical equations (\ref{dyn_cm}) and (\ref{cm_corr}) is:
\begin{align}
\label{fpe}
\frac{\partial}{\partial t}p(x_c, x_r, t) = (L^c_{FP} + L^r_{FP}) p(x_c, x_r, t)
\end{align}
where $L^c_{FP}$ and $L^r_{FP}$ are the Fokker-Planck operators associated with 
the slow $x_c$ and fast $x_r$ degrees of freedom respectively. In the limit of large spring 
constant $k$, the two harmonically coupled particles experience the same potential, 
hence $U(x_c+x_r) \approx U(x_c-x_r)$. As a result, $L^r_{FP} = \frac{\partial}
{\partial x_r}(2kx_r + \frac{D(1-\rho)}{2} \frac{\partial}{\partial x_r})$, which 
admits the Gaussian distribution of mean zero and variance $\sigma^2(x_c, x_r) = 
\frac{D(1-\rho)}{4k}$ as its stationary solution $\psi_0(x_c, x_r)$. Marginalization 
of $p$ using $\psi_0$ leads to the effective drift term for the center of mass motion 
in large $k$ limit and is given by:
\begin{align}
\label{drift}
V'(x_c) = \text{erf}\Big(\frac{x_c+1}{\sigma \sqrt{2}}\Big) - 
\text{erf}\Big(\frac{x_c}{\sigma \sqrt{2}} \Big) + 
\text{erf}\Big(\frac{x_c-1}{\sigma \sqrt{2}} \Big), 
\end{align}
where erf is the error function, and approaches $U'(x_c)$ due to the smallness of 
the variance $\sigma^2$. Hence, in the limit 
of large spring constant the center of mass motion is equivalent to the motion of a single 
particle in the potential given by eqn(\ref{potential}) and with the noise intensity 
modified to $D(1+\rho)$. Such a modification of the noise intensity has strong 
implications on the dynamics of the coupled Brownian particles as shown below. 

The effect of noise correlation on the dynamics of the center of mass can be studied 
using the above result. To investigate the effect, let us calculate the mean first 
passage time $\langle \tau \rangle$ of the center of mass starting at $x_c = -1$ to 
the absorbing boundary at $x_c = 0$. Using the backward Fokker-Planck operator, the 
expression for the mean first passage time reads: 
\begin{align}
\langle \tau \rangle &= (2/D_{\rho}) \int_{-1}^{0} dz ~e^{2U(z)/D_{\rho}}
 \int_{-\infty}^{z} dx ~e^{-2U(x)/D_{\rho}} \nonumber \\
&= D_{\rho} (e^{2/D_{\rho}} - 1) - 1 \nonumber \\
&\approx D_{\rho} e^{2/D_{\rho}}, 
\end{align}
where $D_{\rho} = D(1+\rho)$. 
Hence, the rate of escape of the center of mass from the minima of the potential well 
to the absorbing boundary is $R = 1/\langle \tau \rangle = e^{-2/D_{\rho}}/D_{\rho}$, 
which is of the same form as proposed originally by Kramers. Consequently, it becomes 
nearly impossible for the dimer to escape the potential well for strongly anti-correlated 
noise processes when the coupling between the two monomers is high. 

The strong coupling limit of the dimer motion also allows us to 
calculate the relaxation time of the initial probability density to its steady state. 
We know that in the limit of large spring constant $k$, the dynamics of the center of 
mass follows: $\dot{x}_c = -U'(x_c) + \zeta_c(t)$ admitting $\exp(-U(x_c)/D_{\rho})$ 
as its steady state probability density. Hence, it becomes interesting to 
know the timescale on which the initial density $\delta(x_c+1)$ relaxes towards the 
steady state. In order to calculate the relaxation time $T$, define: 
$Q(t) = \int_{0}^{\infty} dx_c p(x_c, t)$
where $p(x_c, t)$ is the probability distribution associated with the center of mass 
motion and $Q(t)$ is the density of the center of mass being found in the 
basin of attraction of the minima at $x_c = 1$. Using the results in \cite{frisch} 
it is found that: 
$Q(t) - \frac{1}{2} \approx - \frac{1}{2}e^{-\frac{-e^{-2/D_{\rho}}}{D_{\rho}}t} 
- \sqrt{\frac{D_{\rho}}{2\pi}} \frac{e^{-1/D_{\rho}}}{(2/D_{\rho}-1)^2} t^{-3/2}
e^{-t/D_{\rho}}$, 
where the first term is the contribution of the pole of the Laplace transform of 
$p(x_c, t)$ and the second term, which is valid only in the limit of long times is the 
contribution of the branch cut associated with $\hat{p}(x_c, s)$, the Laplace transform 
of $p(x_c, t)$. Using \cite{agudov}, the relaxation time $T$ is given by: 
\begin{align}
T &= \frac{\int_{0}^{\infty} dt (Q(\infty) - Q(t))}{Q(\infty) - Q(0)} \approx 
D_{\rho} e^{2/D_{\rho}}
\end{align}
which is the same as the mean first passage time of the center of mass 
to the absorbing boundary at the peak. It is to be noted that the contribution of the 
branch cut has been ignored in the calculation of the relaxation time $T$, as it is 
valid only in the long-time limit. The dependence of $T$ on noise correlation $\rho$ 
implies that the time to approach stationarity can also be controlled by the 
correlation. 
\begin{figure}
\includegraphics[width=0.5\textwidth]{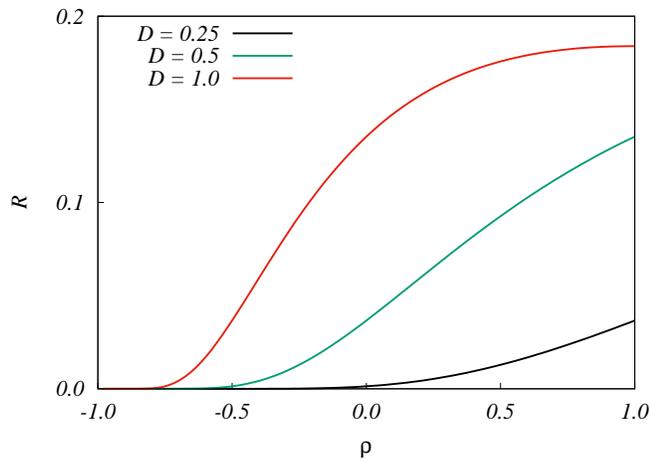}
\caption{Rate of escape $R$ of the dimer from the potential minima to the basin of 
attraction of the other minima as a function of correlation $\rho$ for different values 
of noise intensity $D$ in the strong coupling limit between the monomers. }
\label{fig3}
\end{figure}
Fig.~\ref{fig3} shows the variation of the rate of escape $R$ of the dimer as a function 
of the noise correlation $\rho$ for different values of noise intensity $D$ when the coupling 
between the monomers is very strong. The monotonic variation of $R$ with $\rho$ in the 
strong coupling limit conforms with numerically observed results for relatively weaker values 
of $k$. In addition, as the probability density relaxes towards its stationary value 
at the same rate as the escape rate, the results in Fig.~\ref{fig3} also imply that 
the system may take a very long time to relax to its stationary state when the noise processes 
are strongly anticorrelated. This can leave the dimer confined in the potential minima for 
longer times as compared to independent noise sources and can be employed as a mechanism 
for confinement. With the transient properties of the dimer motion understood, in the 
next section let us generalize to the dynamics of a polymer in a confining potential. 

\section{Generalization to polymer dynamics}
The dynamics of a polymer chain in some potential $U = U(x_1, \ldots, x_N)$ is given 
by the Langevin equation: 
\begin{align}
\label{polymer}
\dot{x}_i = k(x_{i-1}-2x_i+x_{i+1}) - \frac{\partial }{\partial x_i}U({\bf x}) + 
\eta_i(t), 
\end{align}
where $i = 1, \ldots, N$, and the noise processes $\eta_i$ are Gaussian white 
with mean zero and correlations  
$\langle \eta_i(t) \eta_j(t') \rangle  = 2D \rho_{ij} \delta(t-t')$. The diagonal 
elements of the correlation matrix are unity by definition and the off-diagonal 
elements are symmetric and take values from the interval $[-1,1]$, which generalizes 
the dynamics of a polymer chain with independent noise processes \cite{mesfin}. The 
dynamical equations in (\ref{polymer}) can be transformed to the equation for the 
motion of the center of mass and the motion of monomers relative to the center of mass. 
The equation of motion for the center of mass of the polymer: 
\begin{align}
\dot{x}_c = -\sum_i \frac{\partial}{\partial x_i} U({\bf x}) + \zeta_c(t), 
\end{align}
where $\zeta_c = \sum_i \eta_i/N$. The noise process $\zeta_c$ has mean zero and 
correlation $\langle \zeta_c(t) \zeta_c(t') \rangle = \frac{2D}{N^2} (N + 2\sum_{i<j} 
\rho_{ij}) \delta(t-t')$. Now, if the correlations $\rho_{ij}$ are chosen such that 
the term in the brackets becomes small, this can make the polymer to be trapped in 
a metastable state for longer times as compared to the uncorrelated noise processes. 
Positively correlated noise 
processes on the other hand aid in the escape with respect to uncorrelated noises. 
This can be easily understood in the limit when the coupling between the monomers 
is chosen to be very strong, adiabatic elimination of the relative coordinates 
rendering the equation of motion of center of mass: 
$\dot{x}_c = -\tilde{U}(x_c) + \zeta_c(t)$, with $\tilde{U}$ being the effective potential. 
This is equivalent to the dynamics of a single particle in the potential $\tilde{U}$ 
and the thermal degrees of freedom controlled by the parameters $D$ and $\rho_{ij}$. As a 
result, the Kramers formula can be used to calculate the rate of escape from a potential 
minima: 
$R \approx \exp(-N\Delta \tilde{U}/D(1 + 2\sum_{i<j}\rho_{ij}/N))$, where $\Delta\tilde{U}$ is 
the height of the potential barrier. The expression generalizes the previously known 
results for $R$ for uncorrelated noise processes \cite{mesfin, park, lee, sebastian} by 
incorporating noise correlations. Now, for a given value of noise intensity 
$D$, the correlations $\rho_{ij}$ can always be chosen such that the term in the brackets: 
$1 + 2\sum_{i<j}\rho_{ij}/N$, becomes small enough to drastically reduce the magnitude 
of thermal fluctuations preventing the polymer to cross the barrier even when the 
assigned value of $D$ is strong enough to drive the barrier crossing process in the 
absence of noise correlations. On the other hand, if the correlations are chosen such 
that $\rho_{ij} > 0$ for all $i, j$, then these enhance the magnitude of the thermal 
fluctuations thereby making the barrier crossing of the polymer more likely in comparison 
to the case with uncorrelated noises. This generalizes the results of the previous 
sections for dimers with 
correlated noises and has implications towards controlling the rates of chemical 
reactions involving polymers by varying the correlation between the noise processes. 

\section{Conclusions}
In summary, the paper discusses the dynamics of harmonically coupled Brownian particles 
in a symmetric, piecewise linear bistable potential under the effect of correlated 
noise processes. The main result of the study is that for a fixed value of noise 
intensity, positively correlated noise processes aid in the escape of the dimer from 
the metastable states whereas anticorrelated noises tend towards confinement provided 
the particles are not moving completely independent of each other. 
This result has significant implications towards the dynamics of polymers 
in potential fields, e.g.- the rates of chemical reactions involving polymers 
can be controlled by varying the noise correlations and if very strongly anti-correlated 
noises are used, then the polymer can be confined in metastable states for longer 
periods of time. Alternatively, correlated noise sources can be employed to 
confine polymers in a metastable state with the amounts of correlation 
controlling the residence times in the confinement. 
The observations also generalize to dynamics of extended objects in 
potentials with multiple minima, e.g.- transport of a dimer/ polymer in a tilted periodic 
potential \cite{reimann}. The effects of noise correlations in such potentials would 
be observed in the variation of current across the potential, with the current reducing 
for negative correlations and enhanced for positive correlations. Such generalizations 
of the present results will be taken up in future works.

\end{document}